\documentclass[fleqn,usenatbib]{mnras}

\usepackage{newtxtext,newtxmath}

\usepackage[T1]{fontenc}

\DeclareRobustCommand{\VAN}[3]{#2}
\let\VANthebibliography\thebibliography
\def\thebibliography{\DeclareRobustCommand{\VAN}[3]{##3}\VANthebibliography}

\usepackage{graphicx}	
\usepackage{amsmath}	

\title[Overdensities in the EoR]{The Prevalence of Galaxy Overdensities Around UV-Luminous Lyman $\mathbf{\alpha}$ Emitters in the Epoch of Reionization}

\author[E. Leonova et al.]{E. Leonova$^{1,2,3}$\thanks{E-mail: e.leonova@uva.nl; pascal.oesch@unige.ch},
P. A. Oesch$^{1,4}$,
Y. Qin$^{5,6,7}$,
R. P. Naidu$^8$,
J. S. B. Wyithe$^{5,6}$,
S. de Barros$^1$,
R. J. Bouwens$^{9}$,\newauthor
R. S. Ellis$^{10}$,
R. M. Endsley$^{11}$,
A. Hutter$^{12}$,
G. D. Illingworth$^{13}$,
J. Kerutt$^1$,
I. Labb\'{e}$^{14}$,
N. Laporte$^{15,16}$,\newauthor
D. Magee$^{13}$,
S. J. Mutch$^{5,6}$,
G. W. Roberts-Borsani$^{17}$,
R. Smit$^{18}$,
D. P. Stark$^{11}$,
M. Stefanon$^{9}$,\newauthor
S. Tacchella$^{19}$,
A. Zitrin$^{20}$
\\
$^{1}$Department of Astronomy, University of Geneva, Chemin Pegasi 51, 1290 Versoix, Switzerland\\
$^{2}$GRAPPA, Anton Pannekoek Institute for Astronomy and Institute of High-Energy Physics,\\ University of Amsterdam, Science Park 904, 1098 XH Amsterdam, The Netherlands\\
$^{3}$Nikhef, Science Park 105, 1098 XG Amsterdam, The Netherlands\\
$^{4}$Cosmic Dawn Center (DAWN), Niels Bohr Institute, University of Copenhagen, Jagtvej 128, K\o benhavn N, DK-2200, Denmark\\
$^{5}$School of Physics, University of Melbourne, Parkville, VIC 3010, Australia\\
$^{6}$ARC Centre of Excellence for All Sky Astrophysics in 3 Dimensions (ASTRO 3D)\\
$^{7}$Scuola Normale Superiore, Piazza dei Cavalieri 7, I-56126 Pisa, Italy\\
$^{8}$Center for Astrophysics $|$ Harvard \& Smithsonian, 60 Garden Street, Cambridge, MA 02138, USA\\
$^{9}$Leiden Observatory, Leiden University, PO Box 9500, 2300 RA Leiden, The Netherlands\\
$^{10}$Department of Physics and Astronomy, University College London, Gower Street, London WC1E 6BT, UK\\
$^{11}$Steward Observatory, University of Arizona, 933 N Cherry Ave, Tucson, AZ 85721, USA\\
$^{12}$Kapteyn Astronomical Institute, University of Groningen, PO Box 800, NL-9700 AV Groningen, the Netherlands\\
$^{13}$Department of Astronomy and Astrophysics, UCO/Lick Observatory, University of California, Santa Cruz, CA 95064, USA\\
$^{14}$Centre for Astrophysics \& Supercomputing, Swinburne University of Technology, PO Box 218, Hawthorn, VIC 3112, Australia\\
$^{15}$Kavli Institute for Cosmology, University of Cambridge, Madingley Road, Cambridge CB3 0HA, UK\\
$^{16}$Cavendish Laboratory, University of Cambridge, 19 JJ Thomson Avenue, Cambridge CB3 0HE, UK\\
$^{17}$Department of Physics and Astronomy, University of California, Los Angeles, 430 Portola Plaza, Los Angeles, CA 90095, USA\\
$^{18}$Astrophysics Research Institute, Liverpool John Moores University, 146 Brownlow Hill, Liverpool L3 5RF, UK\\
$^{19}$Department of Physics, Ulsan National Institute of Science and Technology (UNIST), Ulsan 44919, Republic of Korea\\
$^{20}$Physics Department, Ben-Gurion University of the Negev, P.O. Box 653, Be'er-Sheva 84105, Israel
}

\date{Accepted XXX. Received YYY; in original form ZZZ}

\pubyear{2021}

\begin{document}
\label{firstpage}
\pagerange{\pageref{firstpage}--\pageref{lastpage}}
\maketitle

\begin{abstract}
Before the end of the epoch of reionization, the Hydrogen in the Universe was predominantly neutral. This leads to a strong attenuation of Ly$\alpha$ lines of $z\gtrsim6$ galaxies in the intergalactic medium.
Nevertheless, Ly$\alpha$ has been detected up to very high redshifts ($z\sim9$) for several especially UV luminous galaxies. Here, we test to what extent the galaxy's local environment might impact the Ly$\alpha$ transmission of such sources. 
We present an analysis of dedicated Hubble Space Telescope (HST) imaging in the CANDELS/EGS field to search for fainter neighbours around three of the most UV luminous and most distant spectroscopically confirmed Ly$\alpha$ emitters: EGS-zs8-1, EGS-zs8-2 and EGSY-z8p7 at $z_\mathrm{spec}=7.73$, 7.48, and 8.68, respectively.  
We combine the multi-wavelength HST imaging with Spitzer data to reliably select $z\sim7-9$ galaxies around the central, UV-luminous sources. 
In all cases, we find a clear enhancement of neighbouring galaxies compared to the expected number in a blank field (by a factor $\sim 3-9\times$). Our analysis thus reveals ubiquitous overdensities around luminous Ly$\alpha$ emitting sources in the heart of the cosmic reionization epoch. We show that our results are in excellent agreement with expectations from the \textsc{Dragons} simulation, confirming the theoretical prediction that the first ionized bubbles preferentially formed in overdense regions. JWST follow-up observations of the neighbouring galaxies identified here will be needed to confirm their physical association and to map out the ionized regions produced by these sources.

\end{abstract}




\section{Introduction}

The Epoch of Reionization (EoR) remains one of the key frontiers of extragalactic studies. Understanding exactly when and how reionization occurred within the first Gyr of cosmic history is a major goal of astronomy over the next decade. While great progress has been made to constrain the overall time frame of reionization through different measurements, including the polarization of the Cosmic Microwave Background \citep[CMB;][]{Planck2018}, the onset and duration of the EoR are still uncertain (e.g., \citealt{Greig17,Naidu20,Bosman21}; also see \citealt{Robertson21} for a recent review).

An efficient tool used to further constrain the evolution of the neutral fraction at $z>6$ is the Ly$\alpha$ line emitted by galaxies (so-called Ly$\alpha$ Emitters; LAEs).
Due to its resonant nature, Ly$\alpha$ is easily scattered and absorbed in the neutral hydrogen in the IGM making LAEs a sensitive probe of the progress of cosmic reionization  \citep[e.g.,][]{MiraldaEscude98,McQuinn2007,Dayal11,Mason2018}. Indeed, the fraction of continuum selected galaxies that do show Ly$\alpha$ has been shown to decrease rapidly when entering the neutral phase of the Universe at $z>6$ (e.g., \citealt{Stark2010,Fontana2010,Schenker2012,Treu13,Pentericci2014,Hoag2019}; also see \citealt{Ouchi20} for a recent review). 

In stark contrast to this overall decrease of LAE fractions in the EoR are the successful detections and spectroscopic confirmations of particularly luminous, continuum-selected galaxies at $z>7$. 
From a UV-luminous sample with $M_\mathrm{UV}\sim-22$ mag identified in the CANDELS fields \citep{Borsani2016}, all four galaxies were subsequently confirmed spectroscopically through their Ly$\alpha$ line \citep{Oesch2015,Zitrin2015,Borsani2016,Stark2017,Pentericci18}, and in some cases also through other UV emission lines. These galaxies thus first hinted at a trend of larger Ly$\alpha$ fractions in more luminous galaxies at $z>7$ compared to fainter sources \citep[see also][]{Ono2012,Mason2018}. This has subsequently been confirmed with larger statistics \citep[e.g.,][]{Jung2020,Endsley2021,Laporte21}, albeit not all luminous sources at $z>7$ revealed detectable Ly$\alpha$.

Several possible explanations for larger detection rates of Ly$\alpha$ in luminous sources have been discussed in the literature. In particular, the sample from \citet{Borsani2016} was initially selected through an IRAC excess, which originates from extremely strong [OIII]+H$\beta$ emission lines, with rest-frame equivalent widths in excess of 800~\AA. Such strong lines seem to be typical at these redshifts \citep[e.g.,][]{Labbe2013,Smit15,deBarros19,Endsley21_oiii}. The strong radiation field required to power such extreme lines might also impact the local environment around these galaxies, increasing the transmission of Ly$\alpha$ \citep[][]{Stark2017,Endsley2021}. An additional contribution of ionizing photons could come from an active galactic nucleus (AGN), which might be present in some luminous sources \citep[][]{Tilvi2016,Laporte17,Mainali18,Endsley2021}.  Additionally, luminous sources have been observed to show significant velocity offsets between their Ly$\alpha$ lines and the systemic redshifts, likely due to strong outflows \citep[e.g.,][]{Erb14,Willott15,Stark2017,Hashimoto2019,Matthee20}. This would shift the red wing of the Ly$\alpha$ lines further away from the IGM damping wing, further increasing transmission. However, it is still not clear whether this effect is large enough to explain the observed luminosity-dependent evolution of the Ly$\alpha$ fraction over $z=6-8$ \citep[e.g.,][]{Mason2018}.

Another possible explanation for a very high Ly$\alpha$ fraction in luminous sources could be that they are embedded in the most overdense regions. These overdense volumes are expected to reionize first thanks to the combined emission from both the luminous galaxies and the numerous fainter sources around them \citep[e.g.,][]{Barkana2004, Dayal2009, Pentericci2011,Endsley2021}. Indeed, the bright galaxies themselves are most likely unable to emit enough ionizing photons to produce an ionized bubble large enough to allow Ly$\alpha$ photons to escape \citep{Wyithe2005}.

Observationally, several galaxy overdensities have now been identified during the EoR both for continuum and narrow-band selected galaxies \citep[e.g.,][]{Trenti12,Ishigaki2016,Hu21}. In several cases, there is strong evidence for a connection between galaxy overdensities and overlapping ionized bubbles that allow Ly$\alpha$ to be transmitted  \citep[e.g.,][]{Castellano2016,Tilvi2020,Endsley2021}. 
In this scenario, reionization  starts in the first overdense regions, implying large spatial fluctuations in the ionized fraction and an inhomogeneous topology. This is indeed what is found in cosmological simulations \citep[e.g.,][]{Iliev06,Hutter17,Geil2017,Kulkarni19, Gronke2020, Ocvirk2020,Qin2021,Smith21}.

In this paper, we extend the search for overdensities and possible ionized bubbles to very high redshifts. In particular, we obtained dedicated HST imaging around three of the most distant and UV-luminous Ly$\alpha$-detected galaxies, in order to search for fainter neighbouring galaxies. In total, we study the environment of four galaxies in the EGS field: three confirmed Ly$\alpha$ emitting sources at $z=7.5-8.7$, in addition to a photometrically selected galaxy at $z\sim9$ (see Sect. \ref{sec:data}).

The paper is organized as follows: in Section \ref{sec:data}, we present the HST and ancillary datasets used. Section \ref{sec:sample} focuses on the selection of our galaxy sample, before we quantify the environment around the UV luminous targets in Section \ref{sec:results} and compare our results with simulations in Section \ref{sec:simcompare}. Finally, we discuss implications for the Lyman-alpha visibility of luminous reionization-era galaxies in Section \ref{sec:conclusions}.

Throughout this paper, we adopt a standard cosmology with $\Omega_{m}=0.3$, $\Omega_{\Lambda}=0.7$, $\mathrm{h}=0.7$. Magnitudes are given in the AB system.

\begin{figure*}
\centering
\includegraphics[width=1.\linewidth]{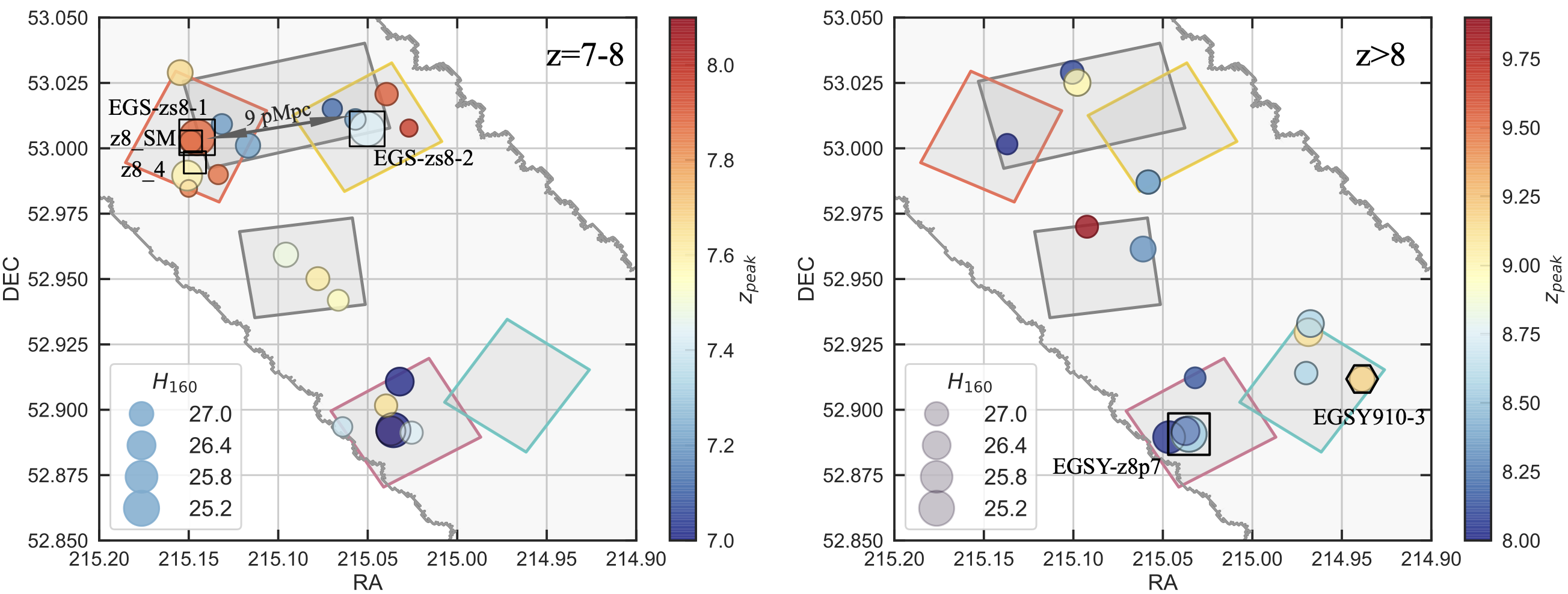}

\vspace{-6pt}
\caption{Position of all selected sources in the CANDLES/EGS field centred around our target galaxies. Sources with redshift $7<z_\mathrm{phot}<8$ are shown in the left panel, while galaxies with $z_\mathrm{phot}>8$ are indicated on the right. The fields are shown North up, East left. The light shaded region shows the CANDELS $H_{160}$ coverage. The darker grey areas show fields where F105W imaging is available and that have been included in our search. The three pointings from our own program were centred on EGS-zs8-1 (orange outline), EGS-zs8-2 (yellow) and EGSY-z8p7 (pink). Also highlighted (in green) is the F105W outline around EGS910-3 presented in \citet{Bouwens2016d} that we include in our analysis.
Black squares correspond to the location of Ly$\alpha$ emitters. The colourbars indicate the photometric redshifts of the sources. The sizes of the circles represent their $H_{160}$ magnitude. Also indicated is the 3D distance of 9 physical Mpc between the two Ly$\alpha$ galaxies at $z\sim7.6$ on the left.}
\label{fig:fieldLayout}
\end{figure*}

\section{Imaging data}
\label{sec:data}

\subsection{Ancillary Data}
The EGS field is one of the five extragalactic fields observed with the Cosmic Assembly Near-infrared Deep Extragalactic Legacy Survey (CANDELS) program \citep{Koekemoer2011, Grogin2011}. It is centred at (RA, DEC)=(214.825000, + 52.825000) and was originally covered with 2-orbit exposures of Advanced Camera for Surveys (ACS) $V_{606}$, $I_{814}$ and Wide Field Camera 3 (WFC3/IR) $J_{125}$ and $H_{160}$ imaging (0.67 orbits in F125W, and 1.33 in F160W). The 3D-HST survey covers the field with $JH_{140}$ (F140W) imaging \citep[][]{Skelton2014,Momcheva2016}, and a few individual follow-up programs have added targeted $Y_{105}$ (F105W) images for a few sources \citep[e.g.,][]{Bouwens2016d}. Additionally, the field has extensive multi-wavelength data from ground-based facilities and was covered with Spitzer's Infrared Array Camera (IRAC) as part of the S-CANDELS program \citep[][]{Ashby15}. The total observed area of the EGS field is $\sim$200 arcmin$^2$. For more information on the ancillary multi-wavelength data in this field see, e.g., \citet{Skelton2014} and \citet{Stefanon2017}.

\subsection{Dedicated Follow-up of Luminous High-Redshift Sources}
\label{sec:folowupdata}

The EGS field is especially rich in early galaxies \citep[e.g.,][]{Bouwens2015}. It has revealed several particularly luminous high-redshift candidates at $z>7$ \citep[e.g.][]{Bouwens2015,Bouwens2019,Borsani2016,Finkelstein2021}, three of which have subsequently been confirmed through Ly$\alpha$ emission lines. These are EGS-zs8-1 at $z_\mathrm{spec}= 7.7302\pm 0.0006$ (\citealt{Oesch2015}; see also \citealt{Stark2017}; \citealt{Tilvi2020}), EGS-zs8-2 at $z_\mathrm{spec} =  7.4770 \pm 0.0008$ \citep{Borsani2016,Stark2017}, and EGSY-z8p7 at $z_\mathrm{spec} = 8.683^{+0.001}_{-0.004}$ (\citealt{Zitrin2015}; see also \citealt{Mainali18}). 
All of these sources are expected to exhibit strong radiation fields due to high-equivalent width [OIII]+H$\beta$ emission lines identified through their Spitzer/IRAC photometry \cite[][]{Borsani2016}. Indeed, highly excited UV lines have been confirmed in the MOSFIRE spectra of EGS-zs8-1 (\ion{C}{iii}; \citealt{Stark2017}) and EGSY-z8p7 (\ion{N}{v}; \citealt{Mainali18}). Especially the detection of the \ion{N}{v} line in EGSY-z8p7 points to a non-negligible, possible contribution from an AGN.

While other particularly distant EoR galaxies have been spectroscopically confirmed since the discovery of these sources in the EGS field \citep[e.g.][]{Oesch2016,Jiang21,Hashimoto2018,Jung2020,Laporte21}, they still rank among the most distant detected Ly$\alpha$ emission lines \citep[see also Larson et al., in prep;][]{Larson2020}. For more information on the luminous Ly$\alpha$ sources, see Section \ref{sec:maintargets} and Tables \ref{table1} and \ref{table2}.

To explore whether these very early Lyman-alpha emitters commonly reside in overdense regions, we have obtained dedicated, additional multi-colour HST WFC3/IR observations around these three luminous galaxies \citep[GO-15103;][]{HSTproposal}. In particular, we added the missing F105W filter that is necessary to reliably identify $z\sim7.5-9$ galaxies and measure improved redshifts, in addition to deeper F125W and F160W imaging. The program was designed to identify galaxies that are up to an order of magnitude less luminous than the main targets, which have $H_{AB}=25.0-25.3$ mag.
In particular, we obtained 5-orbit exposures in three individual pointings centred on each of the three luminous LAEs (2 orbits in F105W, 1.5 in F125W, and 1.5 in F160W). When combined with the previous, CANDELS HST imaging our data reach 5$\sigma$ limits of 27.9 mag in F105W, 27.5 mag in F125W, and 27.5 mag in F160W, respectively.

In addition to our dedicated HST data, we retrieved all the HST ACS and WFC3/IR data available around the EGS sources from the HST archive and combined them into one mosaic as part of the Hubble Legacy Field project (Illingworth et al, in prep). 

\subsection{Source Detection and Multi-Wavelength Photometry}
\label{sec:catalog}

The HST images in different bands were convolved to the same point-spread function (PSF; using convolution kernels analogous to the ones made available by the 3D-HST team; \citealt{Skelton2014}).
We then use SExtractor \citep{Bertin1996} to detect sources in a combined F125W+F160W image and measure their colours through matched, circular apertures. In particular, we use apertures with 0\farcs5 diameter to measure colours, which are then corrected to total fluxes using the SExtractor AUTO fluxes in the F160W image.

In order to obtain IRAC photometry in all four channels, we follow the same procedure as described in \citet{Stefanon21}. To overcome confusion, we subtract contaminating flux of neighbouring galaxies using the \texttt{Mophongo} tool \citep[see, e.g.,][]{Labbe2013,Labbe2015}, before measuring the IRAC fluxes in 1\farcs8 diameter apertures and correcting to total fluxes. When the contamination by neighbour is too large for a reliable subtraction, the IRAC fluxes are flagged and ignored in the subsequent spectral energy distribution fits. 
We use the same procedure to derive photometry from ground-based K-band images. Overall, we thus obtain 11-band spectral energy distributions (SED) for all sources (6$\times$ HST, K-band, and 4$\times$ IRAC), spanning 0.6 to 8 \micron.

\section{Sample Selection }
\label{sec:sample}

\subsection{\bf{Central LAE Targets}} \label{sec:maintargets}

Our HST follow-up observations are centred around three UV luminous galaxies at $z\geq 7.5$ that have been confirmed through Ly$\alpha$ in the EGS field (see Section \ref{sec:folowupdata}). One of these, EGS-zs8-1, has subsequently been observed through narrow-band observations, from which two fainter LAEs have been identified in its immediate proximity \citep{Tilvi2020}. These latter sources have also been confirmed through Ly$\alpha$ spectroscopy and they were named z8\_SM at $z=7.767$ and z8\_4 at $z=7.748$. They likely sit in a common, ionized bubble with EGS-zs8-1, since they are only separated by 0.7 to 0.8 physical Mpc (pMpc) from each other and their estimated ionized bubble radii reach $0.5-1.0$ pMpc.

Additionally, the galaxy group around EGS-zs8-1 is separated by only 9.1 pMpc from the other UV-luminous, spectroscopically confirmed source EGS-zs8-2 at z=7.48. This distance is smaller than the average bubble radius of  $\lesssim$10 pMpc expected at the end of cosmic reionization \citep{Wyithe2005,Geil2017}, and points at a larger scale overdensity in this field at this redshift.
One of the main goals of our follow-up observations was to further characterize this overdensity through fainter sources. Hence, we centred two of our HST pointings around EGS-zs8-1 and EGS-zs8-2, respectively (orange and yellow outlines in Fig \ref{fig:fieldLayout}).

The last pointing was centred around EGSY-z8p7, one of the most distant known Ly$\alpha$ emitters with $z_\mathrm{spec}=8.683$ (pink outline in Fig \ref{fig:fieldLayout}). This source has a potential companion at $<10$ pMpc: the galaxy EGS910-3 (\citealt{Bouwens2016d}, see also \citealt{Bouwens2019}). On the sky, these two sources are separated by only 3.7 arcmin. However, the source only has a photometric redshift measurement ($z_\mathrm{phot}=9.0^{+0.5}_{-0.7}$ as derived in \citealt{Bouwens2016d}) and has not been confirmed through spectroscopy so far. Hence, the actual physical separation from EGSY-z8p7 is uncertain. 
The source EGS910-3 has also been targeted by F105W follow-up observations, which we retrieve from the archive and include in our analysis (see, e.g., Fig \ref{fig:sed}).

\subsection{Selection of Neighbouring z=7-9 Candidates}

\subsubsection{Selection Criteria} 

The main goal of our program was to identify fainter sources around the central UV-luminous targets. We thus use the HST-detected multi-wavelength catalogs described in Section \ref{sec:catalog} to select candidate high-redshift sources.
Specifically, we use a combination of detection and non-detection criteria as expected for Lyman break galaxies together with a photometric redshift selection. We only consider sources that are significantly detected in both the F125W and F160W bands with the following criteria:

   \begin{equation}
       S/N(J_{125})>3 \land (S/N)_{\mathrm{det}}>5 \land S/N(H_{160})>4,
       \label{det}
   \end{equation}
where $(S/N)_{\mathrm{det}}$ corresponds to the signal-to-noise in the combined F125W and F160W image. These criteria allow us to identify sources down to $H_{160}=27.7$ mag.
Additionally, the sources were required to not be detected in the two optical HST filters with: $S/N(V_{606})<2 \land S/N(I_{814})<2$. In selecting these sources, we have only included areas that also have F105W imaging coverage, either from our own follow-up program or from archival observations (darker areas in Fig \ref{fig:fieldLayout}). 

Photometric redshifts are then derived for all the sources that pass the above criteria. We use the python version of the publicly available code \texttt{EAZY}\footnote{\url{https://github.com/gbrammer/eazy-py}} \citep{Brammer2008} which provides the photometric redshift probability distribution function (pdf), $\mathcal{P}(z)$, based on a $\chi^2$ fit of SED models to the observed photometry. Specifically, we use the template set `eazy\_v1.1' which includes emission lines. We have also tested other template sets, but have not found large differences. \texttt{EAZY} also accounts for the intergalactic absorption by neutral hydrogen \citep{Madau1995}. The redshifts have been searched over $0.1 \leq z \leq 12$.

Based on the \texttt{EAZY} output we then only include sources that have a best-fit photometric redshift in our range of interest, depending on the central target sources. For the fields around EGS-zs8-1 and -2, we select galaxies with $z_\mathrm{phot}=7-8$. To ensure that these redshifts are reasonably accurate, we exclude galaxies with a photometric redshift pdf that is too wide, i.e., galaxies are required to have $\mathcal{P}(6 \leq z \leq 9) \geq 50\%$. 
For the fields around EGSY-z8p7, we select sources with $z_\mathrm{phot}>8$ that have $\mathcal{P}(7 \leq z \leq 10) \geq 50\%$.
In Figure \ref{fig:sed}, we illustrate the SEDs and redshift pdfs of the central UV-luminous Ly$\alpha$ emitters and EGS910-3. As can be seen, the photometric redshift determinations of the Ly$\alpha$ sources are in excellent agreement with their spectroscopic redshifts, with a very narrow $\mathcal{P}(z)$. For the last source, we confirm the photometric redshift solution from \citet{Bouwens2016d}, as we measure a consistent value of $z_\mathrm{phot}=9.18^{+0.95}_{-0.55}$.

\subsubsection{Exclusion of Dwarf Stars}

Cool dwarf stars can mimic the colours of high-redshift Lyman break galaxies \citep[e.g.][]{Wilkins14}. They can remain undetected in optical bands (in F606W and F814W filters) and peak in the Near-infrared (NIR).
To remove such sources from our catalogs of LBG candidates we use a combination of two different measurements. First, we identify potential unresolved sources using the SExtractor parameter $\mathrm{CLASS\_STAR}$, and second we run \texttt{EAZY} with a set of dwarf star templates fixed at $z=0$ to compare the $\chi^2$ from both dwarf stars and galaxy templates.
The $\mathrm{CLASS\_STAR}$ parameter on its own is not reliable, especially at faint magnitudes.

Objects with $\mathrm{CLASS\_STAR}$ more than 0.6 and $\chi^2$ from star templates less than $\chi^2$  from galaxy template are removed from the catalogs:
\begin{equation}
\centering
   \mathrm{CLASS\_STAR}>0.6 \land (\chi^2_{\mathrm{galaxy}}/\chi^2_{\mathrm{star}})>1
\end{equation}

In total, only two sources have been removed based on these criteria. Visual inspection of these sources indeed confirmed them to be unresolved, and hence more likely to be stars than high-redshift galaxies.

\begin{table*}
\caption{Table with parameters of all sources with redshift $z\sim7-8$ identified in the F105W footprints around the two LAEs.}\label{table1}

\begin{tabular}{ccccccccc}
\hline\hline

 ID & RA & DEC & H$_{160}$ [mag] & z$_\mathrm{phot}$ &  $\int\mathcal{P}(6\leq z\leq9)$ & z$_\mathrm{spec}$  &
 Separation$^a$ [$'$] & References\\\hline\hline\\[-5pt]

\multicolumn{9}{c}{\textbf{EGS-zs8-2 group}}\\[5pt]

    EGS-zs8-2 &     215.0503352     &  53.0074442  &   $25.25 \pm 0.03$   &          $7.43^{+0.08}_{-0.09}$    & 1.00 & $7.477$ & 0.0 &(1),(3) \\
EGSY-94562 & 215.0394433 & 53.0207175 & $27.2 \pm 0.3$ & $7.90^{+0.04}_{-1.90}$ &0.55 & $-$ & 0.9 &\\
EGSY-80113 & 215.056893 & 53.0110575 & $27.4 \pm 0.2$ & $7.22^{+0.02}_{-1.22}$ &0.50 & $-$  & 0.3 &\\
EGSY-76521 & 215.0697905 & 53.0151445 & $27.5 \pm 0.2$ & $7.12^{+0.20}_{-0.93}$ & 0.90 &  $-$ & 0.8 &\\
EGSY-91929 & 215.0269545 & 53.0077343 & $27.7 \pm 0.2$ & $7.96^{+0.20}_{-0.95}$ &0.93 & $-$ & 0.8 &\\

\hline\hline\\[-5pt]

\multicolumn{9}{c}{\textbf{EGS-zs8-1 group}}\\[5pt]

EGS-zs8-1 & 215.1453377 & 53.0042107 & $25.12 \pm 0.04$ & $7.88^{+0.06}_{-0.05}$  &1.00 & 7.728 & 0.0 &  (1),(2),(3),(4)\\
EGSY-23959 & 215.1508931 & 52.9895618 & $26.1 \pm 0.1$ & $7.59^{+0.13}_{-0.36}$ &1.00 & $-$ & 0.9 &\\
EGSY-46810 & 215.154764 & 53.0289543 & $26.8 \pm 0.1$ & $7.67^{+0.23}_{-1.67}$ &0.66 & $-$ & 1.5 &\\
EGSY-46009 & 215.1168772 & 53.0010665 & $26.9 \pm 0.2$ & $7.22^{+0.10}_{-1.16}$ &0.86 & $-$ & 1.0 &\\
z8\_SM & 215.1487577 & 53.0025984 & $27.2 \pm 0.1$ & $7.89^{+0.16}_{-0.29}$ &1.00 & 7.767 & 0.2 & (4)\\
EGSY-44464 & 215.1314765 & 53.0092156 & $27.4 \pm 0.2$ & $7.21^{+0.09}_{-1.21}$ & 0.76 & $-$ & 0.6 &\\
EGSY-31530 & 215.1334993 & 52.9899563 & $27.5 \pm 0.2$ & $7.86^{+0.41}_{-1.86}$  & 0.51 & $-$ & 1.0 &\\
EGSY-21312 & 215.150041 & 52.9846593 & $27.7 \pm 0.2$ & $7.85^{+0.24}_{-0.70}$ &0.96 & $-$  & 1.2 &\\

\hline
z8\_4$^*$ & 215.1465376 & 52.994614 & $26.8 \pm 0.3$ & $6.16^{+0.90}_{-0.48}$ & $-$ & 7.748 & 0.6 & (4)\\

\hline\hline
\end{tabular}

\vspace{1ex}
\footnotesize{\raggedright Notes: $^a$ Separation from the central luminous source in arcmin. \par}
\footnotesize{\raggedright $^*$z8\_4 was not selected in our sample due to a lower photometric redshift. \par}
\footnotesize{\raggedright References: (1) \citet{Borsani2016}, (2) \citet{Oesch2015}, (3) \citet{Stark2017}, (4) \citet{Tilvi2020} \par}
\end{table*}


\begin{table*}
\centering
\caption{Table with parameters of all sources with redshift $z\sim8-9$ identified in the F105W footprints around the two $z>8$ galaxies.}\label{table2}

\begin{tabular}{ccccccccc}
\hline\hline

 ID & RA & DEC & H$_{160}$ [mag] & z$_\mathrm{phot}$ & $\int\mathcal{P}(7\leq z\leq10)$ & z$_\mathrm{spec}$ & Separation$^a$ [$'$]& References
\\\hline\hline\\[-5pt]

\multicolumn{9}{c}{\textbf{EGSY-z8p7 group}}\\[5pt]

EGSY-z8p7 & 215.0354269 & 52.8906918 & $25.19 \pm 0.03$ & $8.55^{+0.14}_{-0.15}$  &1.00 & $8.683$ & 0.0 & (1),(2),(3),(4)\\
EGSY-6887$^*$ & 215.0465028 & 52.8895178 & $25.8 \pm 0.1$ & $8.05^{+0.22}_{-0.54}$ & 0.95 & $-$ & 0.4 & \\
EGSY-12580$^\dagger$ & 215.0373571 & 52.8918284 & $26.4 \pm 0.2$ & $8.07^{+0.19}_{-0.29}$  & 0.99  & $-$ & 0.1 &\\
EGSY-27727 & 215.031966 & 52.9122172 & $27.3 \pm 0.2$ & $8.15^{+0.79}_{-1.78}$  &0.68 & $-$ & 1.3 & \\
\hline\hline\\[-5pt]

\multicolumn{9}{c}{\textbf{EGS910-3 group}}\\[5pt]

EGS910-3  & 214.9386673 & 52.91176 & $26.8 \pm 0.2$ & $9.18^{+0.95}_{-0.55}$  &0.63 & $-$ &  0.0 & (2),(3)\\
EGSY-66990 & 214.968693 & 52.9296537 & $26.4 \pm 0.2$ & $9.12^{+0.14}_{-1.08}$  &0.95 & $-$ & 1.5 &\\
EGSY-69674 & 214.9675507 & 52.932962 & $26.6 \pm 0.1$ & $8.60^{+0.25}_{-0.44}$  &0.99 & $-$ & 1.6 & (3),(4) \\ 
EGSY-56680 & 214.9699133 & 52.9140338 & $27.2 \pm 0.1$ & $8.60^{+0.14}_{-2.60}$ &0.66 & $-$ & 1.1 & \\
\hline\hline
\end{tabular}

\vspace{1ex}
\footnotesize{\raggedright Notes: $^a$ Separation from the central luminous source in arcmin. \par}
\footnotesize{\raggedright $^*$ This source lies very close to a clumpy foreground galaxy, which could affect its photometric measurements. \par}
\footnotesize{\raggedright $^\dagger$ This source has a close companion, which was excluded from our list due to its photometric redshift just below the cutoff of our selection.  \par}
\footnotesize{\raggedright References: (1) \citet{Zitrin2015}, (2) \citet{Bouwens2016d}, (3) \citet{Bouwens2019}, (4) \citet{Finkelstein2021}\par}
\end{table*}

\subsubsection{Final Sample} 

After excluding two likely dwarf stars, our final sample of high-redshift galaxies within the four WFC3/IR pointings around the four UV-luminous central targets consists of 21 sources in total. Their positions are shown in Fig. \ref{fig:fieldLayout}. In the following we will discuss the individual environment around the four central sources.

With 8 sources, the pointing around EGS-zs8-1 revealed the largest number of high-redshift galaxy candidates at $z=7-8$ (including the central source itself). The EGS-zs8-2 field resulted in 5 sources. Note that the spectroscopically confirmed source z8\_4 is not included in our photometric sample due to a photometric redshift that was too low. For completeness, we show the galaxy in Fig. \ref{fig:fieldLayout}.

At higher redshifts, the fields around EGSY-z8p7 and EGS910-3 reveal 4 sources with $z_\mathrm{phot}>8$ each. All these sources are listed in Tables \ref{table1} and \ref{table2}. The photometric redshift distribution functions and stamps for the luminous sources can be found in the appendix in Fig \ref{fig:sed}, while the fainter, neighbouring sources are shown in Fig. \ref{fig:egsz8p7group} to \ref{fig:egszs81group}.

\begin{figure*}
\centering
\includegraphics[width=.48\textwidth]{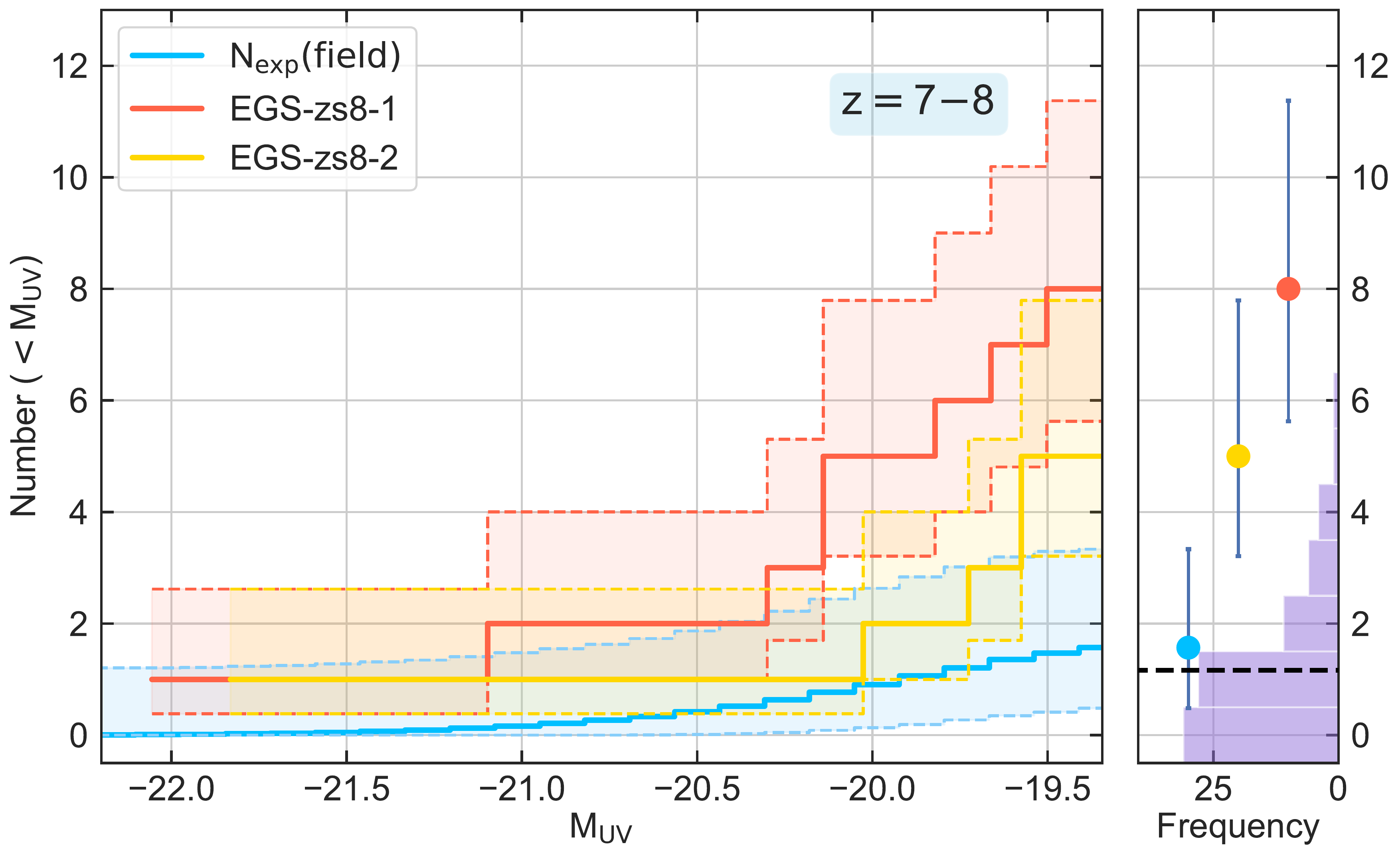}
\medskip
\includegraphics[width=.48\textwidth]{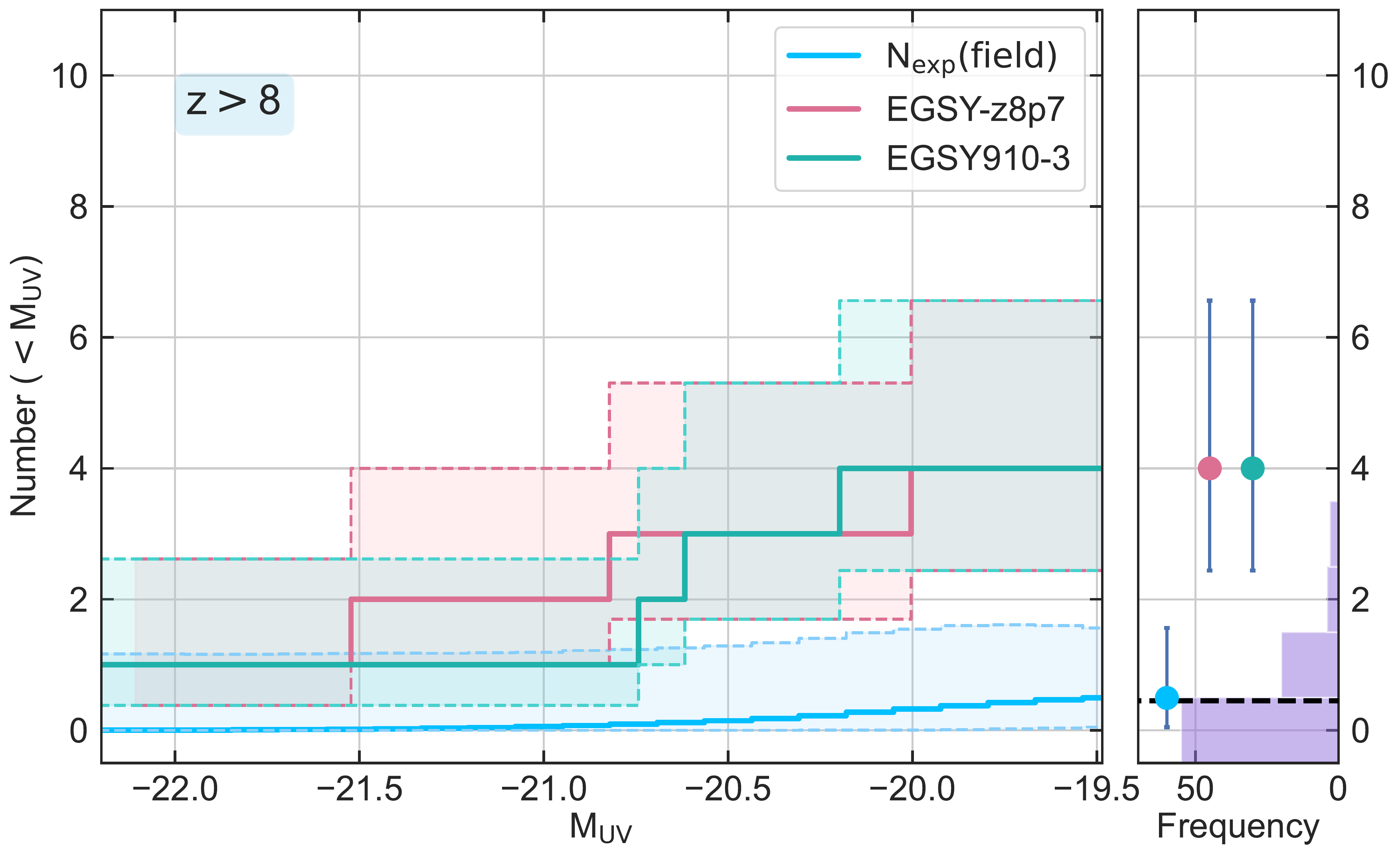}
\caption{The cumulative number of the sources within an area of 4.5 arcmin$^2$ at $z=7-8$ (left) and $z>8$ (right). The left panels show the cumulative numbers as a function of absolute magnitude. The blue line corresponds to the expected number based on the UV LF and our selection function, while the coloured lines correspond to the actual numbers of galaxies that we have identified in the HST data in the different fields (orange for EGS-zs8-1; yellow for EGS-zs8-2;  pink for EGSY-8p7; green for EGS910-3). The shaded regions correspond to the 16th to 84th percentile uncertainties for a Poisson distribution convolved with cosmic variance (see text for details). The right panels show the total number of sources selected down to our magnitude limits using the same colours as for the cumulative histograms,with the points offset horizontally for clarity. Additionally, the purple histograms show the distribution of high-redshift galaxies selected in 82 independent 4.5 arcmin$^{2}$ cells in the GOODS-S and GOODS-N fields, using the same selection criteria as for our HST data in the EGS field (see section \ref{sec:empiricalapproach}). The dashed black line corresponds to the mean in the GOODS fields, which is consistent with the mean expected number from the UV LF (blue). As is evident from these figures, the environment around all four UV luminous sources are overdense relative to the general field. The enhancement ranges from 3$\times$ to 9$\times$.
}
\label{fig:cumnumbfinal}
\end{figure*}

\section{The Environment of UV Luminous Galaxies}
\label{sec:results}

A principle goal of our work is to quantify the environment and possible overdensities of UV luminous sources with Ly$\alpha$ emission at $z>7$. To do this, we first need to estimate the expected number of galaxies in a given area, which we do based on two different approaches that are described in the following subsections.

\subsection{Expected Number of Galaxies}

\subsubsection{Based on the UV Luminosity Function}

The first approach to estimate how many galaxies we expect to find within a given area is based on the UV luminosity function (LF) and the galaxy selection efficiency. Following previous literature \citep[e.g.][]{Castellano2016}, we estimate the selection efficiency of our survey by simulating the photometry for a large number of high-redshift galaxies. Specifically, we introduce $>60,000$ 
artificial galaxies with a range of known (input) magnitudes $24-29$ mag, colour distribution, and redshifts $6<z<10$. 

Since high-redshift galaxy selections can be  sensitive to the UV colours, we simulate the artificial sample with a colour distribution that matches observations. To do this, we start with a \citet{BC03} template with an age of 100 Myr and a constant star-formation history.  \citet{Calzetti00} dust is then applied following a normal distribution of E(B-V) with a mean of 0.1 and a $\sigma$ = 0.15, truncated at zero \citep[following][]{Finkelstein2021}. Nebular continuum and emission lines have been added empirically based on the total number of ionizing photons produced by the underlying SED and using line ratios from \citet{Anders03}. Finally, IGM attenuation is applied according to the prescription of \citet{Madau1995} and set by the simulated redshift of each galaxy. For $z>7$, this essentially corresponds to a sharp break at the redshifted Ly$\alpha$ line.
The final SEDs are then convolved with all the HST and Spitzer IRAC transmission curves to derive the expected, intrinsic fluxes. These are subsequently disturbed using Gaussian flux uncertainties, based on the median flux $\sigma$ as measured for real sources in our HST footprints in the respective filters.

We then perform the same analysis as for the real catalog. First, the detection and optical non-detection criteria are applied and the \texttt{EAZY} photometric redshifts and pdfs are computed as was done for the real galaxies.
This allows us to compute the completeness in bins of input magnitude and redshift as the ratio between the number of galaxies that pass our selection and the number of input galaxies in the respective bin:
\begin{equation}
    {\cal{P}}(M,z)=\frac{N_{\rm{sel}}}{N_{\rm{in}}}.
\end{equation}
Now that we have the selection efficiency, it is straightforward to compute the expected number of galaxy detections based on the empirical  UV LF. To do this, we use the Schechter parameter evolution derived in \citet{Bouwens2021} at $z \sim 2-10$:

\begin{equation}
    \begin{cases}
 M_{\rm{UV}}^*=-21.03 -0.04(z-6) \\
 \phi^*/\rm{Mpc}^{-3}=0.4\times 10^{-3} \times
 10^{-0.33(z-6) -0.24(z-6)^2} \\
 \alpha=-1.94-0.11(z-6) \\
    \end{cases}
    \label{bouwcr}
\end{equation}

The expected number of detections can then be derived by multiplying the UV LF with the probability that a galaxy of absolute magnitude $M$  at redshift $z$ is selected, $\mathcal{P}(M,z)$, within the same area as used for real sources.
 
  \begin{equation}
     N_\mathrm{exp}=\int \Omega \frac{dV_{\rm{com}}}{dz}  dz\int\phi(M) \mathcal{P}(M,z) dM
     \label{eqexp}
 \end{equation}

\noindent where $\phi(M)$ is the UV LF, $\Omega$ is the survey area for each HST field, i.e., 4.5 arcmin$^2$ for a WFC3/IR pointing.

For the $z\sim7-8$ galaxy samples around EGS-zs8-1 and -2, we derive an expected number of 1.57 galaxies per WFC3/IR pointing down to our detection limit of 27.7 mag (see Fig. \ref{fig:cumnumbfinal}). For the $z\sim8-9$ samples, the expected number is only 0.45 sources per WFC3/IR field.

\subsubsection{Empirical Approach}
\label{sec:empiricalapproach}

Another way to determine whether a given HST field might be overdense can be derived empirically using the number counts from another field that is as deep or deeper than the field in question. We can thus use the CANDELS/GOODS-North and CANDELS/GOODS-South fields with an area of $\sim$170 arcmin$^2$ each, from which high redshift galaxy sources have been selected in the past. Specifically, we use the sample from \citet{Bouwens2015}, for which multi-wavelength catalogs are available from GREATS \citep[][]{Stefanon21GREATS}. For consistency, we re-derive the photometric redshifts and pdfs for this sample using the same \texttt{EAZY} setup as for our main selection. 
We then select the groups of sources with best-fit redshifts at $7 \leq z \leq 8$ and $8 \leq z \leq 9$. From these, we limit the sources at $H_{160}=27.7$ mag, which corresponds to the 5$\sigma$ detection limit in our EGS data around the Ly$\alpha$ emitters.

To get a histogram of expected sources in an area equivalent to a single WFC3/IR pointing, we divide the GOODS-S and GOODS-N fields into circular cells with radius 1.2 arcmin, corresponding to an area of 4.5 arcmin$^2$ each, and count the number of the selected sources within each cell. 
The positioning of the cells was chosen on a grid allowing for a small ($<10\%$) overlap between different areas. In total, the GOODS-S field was thus split into 42 independent regions, while the GOODS-N field was divided into 40 separate cells.

The number count histograms of these 82 regions are also shown in Fig. \ref{fig:cumnumbfinal}. The average number of sources per cell is 1.2 at $z\sim7-8$ and 0.45 at $z\sim8-9$, which is consistent with the average expected number from the UV LF derived in the previous section. However, the empirical histograms further allow us to visualize the effect of cosmic variance. 
For example, out of 82 independent cells in both GOODS fields, none contain 8 sources, while only 2 contain $\geq$5 $z\sim7-8$ galaxies. Empirically, finding $\geq$5 galaxies is thus expected only with a 2.4\% probability.

At higher redshift, the largest number of galaxies in a given 4.5 arcmin$^2$ region is 3. Based on this, our finding of 4 (5-8) sources around the UV luminous LAEs at $z>8$ ($z=7-8$) clearly indicates that these galaxies live in early overdensities. 
We quantify this in more detail in the next section.

\begin{figure}
\centering
\includegraphics[width=\linewidth]{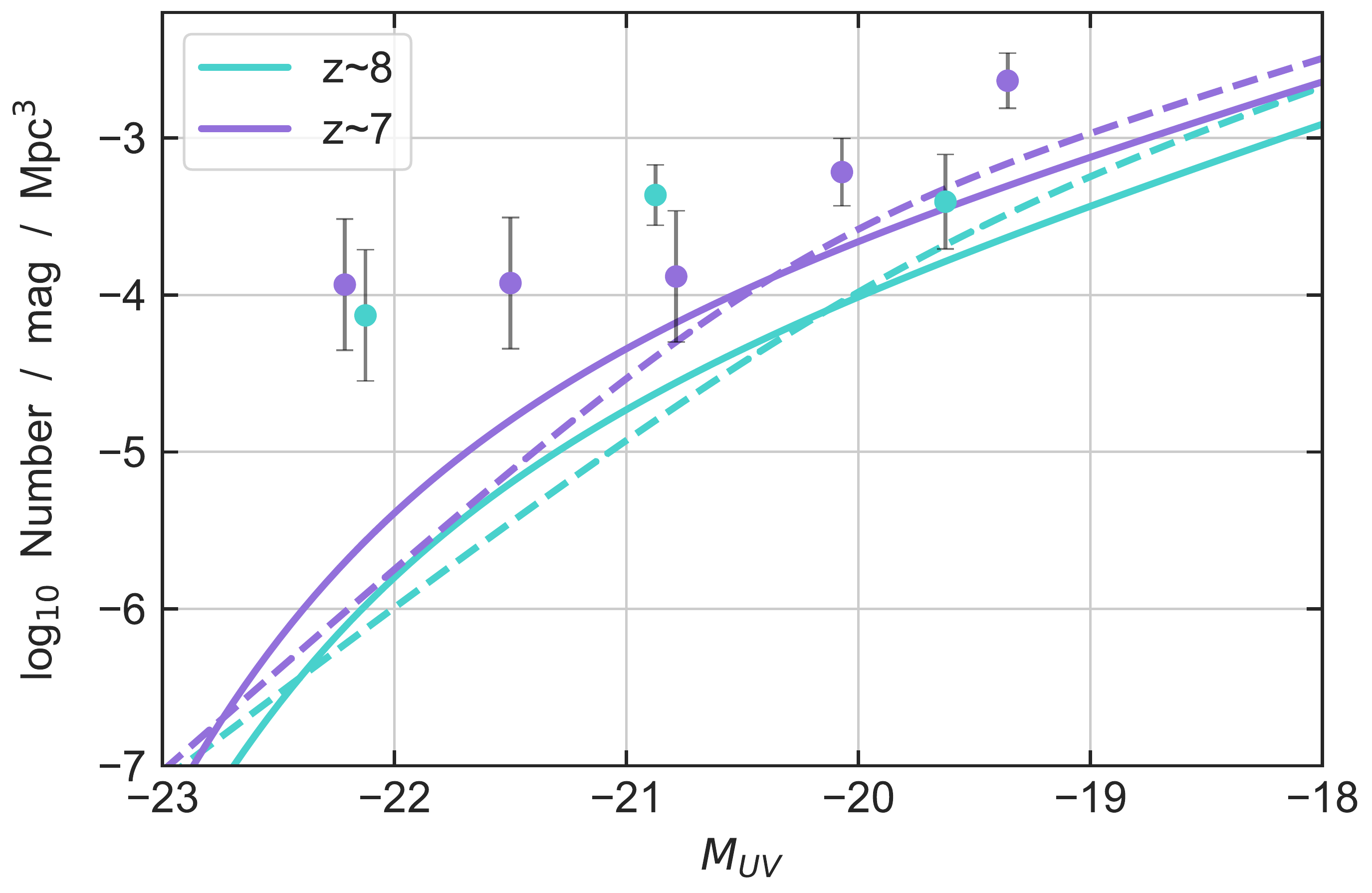}
\vspace{-10pt}
\caption{UV luminosity functions in the EGS field centred around the four UV-luminous sources (dots) compared to expected average Schechter LFs as derived in \citet[][solid lines]{Bouwens2021}, and double power-law LFs \citet[][dashed lines]{Bowler2020}. The purple curve shows the LFs for the $z=7-8$ galaxies, while the cyan line corresponds to $z=8-9$. This analysis shows that these UV luminous sources lie in areas with a general density enhancement of $3-9\times$ that extends to fainter galaxies.
}
\label{fig:UVLF}
\end{figure}

\subsection{Prevalent Overdensities Around UV Luminous Galaxies}

We now ready evaluate the environment of the HST fields around the UV luminous sources. We start with the two Ly$\alpha$ sources at $z\sim7.5$, around which we find 5 and 8 sources in 4.5 arcmin$^2$. This is 3$\times$ and $5\times$ overdense relative to the expected number of 1.57 sources based on the UV LF. 

In order to compute more accurate probabilities to find 5 or 8 sources when 1.57 are expected we specifically account for cosmic variance. To do this, we use the publicly available cosmic variance calculator \texttt{galcv}\footnote{\url{https://github.com/adamtrapp/galcv}} of \citet{Trapp20}. For a 4.5 arcmin$^2$ field and a depth of $\Delta z=1$, this predicts a relative $\sigma_{CV}$ of 32\% at $z=7.5$ and 38\% at $z=8.5$, down to our magnitude limits. These numbers are similar (albeit somewhat smaller) than the predicted cosmic variance from \citet{Trenti2008}. 

The full probabilities are then derived from a Monte Carlo simulation in which we first perturb the expected numbers of galaxies from the UV LF by the relative cosmic variance and then draw from the Poisson distribution. Doing this, we find that the probability to find 8 sources, when 1.57 are expected is indeed small with 0.08\%. This corresponds to a 3.2$\sigma$ result. Similarly, the probability to find $\geq5$ galaxies is only 3.1\%, consistent with our empirical estimate above. This still represents a significance of 1.9$\sigma$.

For the higher redshift fields, we find 4 sources in each pointing, while only 0.46 were expected on average. This corresponds to an enhancement of 8.7$\times$. Doing the same calculation as above, we find that the probability of finding 4 (or more) sources is only 0.2\%, corresponding to a 2.9$\sigma$ result for each field separately.
These calculations show that overdensities around UV luminous sources have been significantly detected in all four cases, with overdensity values ranging between $3-9\times$. 

This excess can also be seen when looking at the UV LF derived in our target fields alone. Fig. \ref{fig:UVLF} shows these UV LFs at $z=7-8$ and $z=8-9$ as derived from the two WFC3/IR pointings centred at the respective redshifts. To compute this, we use the same selection functions as derived in Section 4.1.1 and compute the effective volume in a given magnitude bin. Given the small area (of 9 arcmin$^2$) centred on UV luminous galaxies, the UV LF in these fields is significantly enhanced in the brightest bins (with $M_{UV}\sim-22$). However, as can be seen, at the fainter end these UV LFs continue to lie significantly above the expected field average values. This is consistent with the hypothesis that these UV luminous galaxies lie in overdensities that are generally enhanced relative to the rest of the Universe. Our results thus indicate that the most UV luminous sources can indeed be used to pinpoint the first overdense regions in the early Universe.

\section{Expectations from Simulations}
\label{sec:simcompare}
In this section we discuss what simulations predict for the
environment of UV luminous galaxies similar to those we have
observed.

Specifically, \citet{Qin2021} studied the environment of $z=8$ galaxies in the \textsc{Meraxes} semi-analytical model \citep{Mutch16,Qin2017MNRAS.472.2009Q} as part of the \textsc{Dragons}  simulation program \citep{Poole16}.

\subsection{Number of Neighbouring Galaxies}

\citet{Qin2021} predicted that the most luminous galaxies preferentially lie in overdense regions, in agreement with what we find here from observations. In particular, \citet{Qin2021} used their simulations to compute the number of expected sources within a WFC3/IR field-of-view. This can be directly compared to our observational data.

\begin{figure}
\centering
\includegraphics[width=\linewidth]{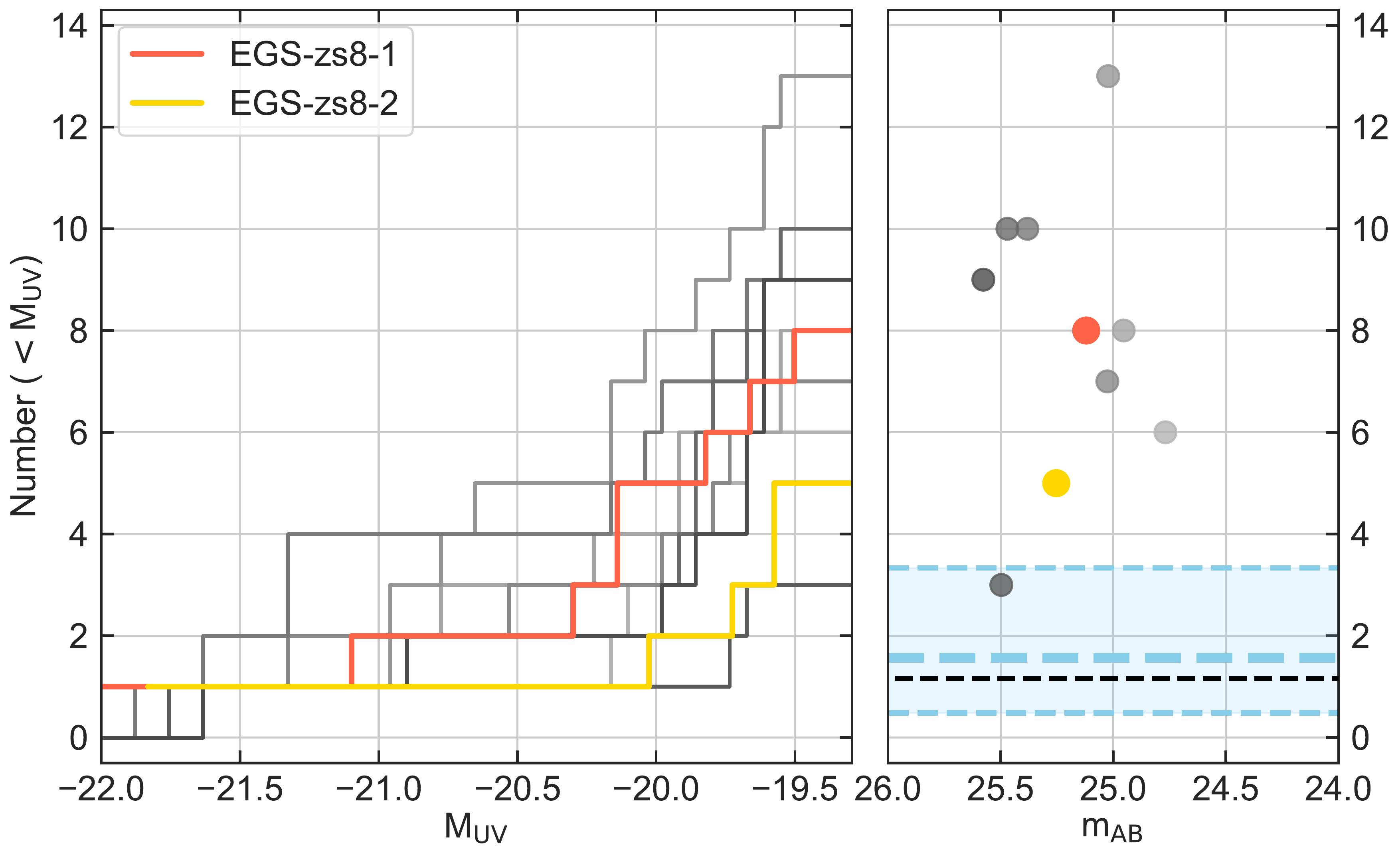}
\vspace{-10pt}
\caption{The cumulative number of the sources within an area of 4.5 arcmin$^2$ at $z\sim 8$, both in simulation and in our observations. The left panel shows the cumulative numbers as a function of absolute magnitude. The coloured lines correspond to the actual numbers of galaxies that we have identified in the HST data in different fields (orange for EGS-zs8-1; yellow for EGS-zs8-2). The grey lines correspond to the numbers from the \textsc{Dragons} simulation (\citealt{Qin2021}). The right panel shows the total number of sources as a function of the magnitude of the brightest source selected down to our 27.7 magnitude limit using the same colours as for the cumulative histograms. The blue line corresponds to the expected number based on the UV LF and our selection function (same as Fig \ref{fig:cumnumbfinal}), with the shaded blue region showing the 16th to 84th percentile. The horizontal black dashed line shows the average number of sources in the GOODS fields within the same area. Both the simulations and the observations thus find a boost of $\sim3-5\times$ in the number of galaxies around the most luminous sources.
}
\label{fig:cumnumbsim}
\end{figure}

Figure \ref{fig:cumnumbsim} shows the cumulative number of galaxies expected around the eight most luminous central sources (having $M_\mathrm{UV}<-21.5$) in the simulations down to the same magnitude limit as used in our observations. As can be seen, the cumulative number varies between 3 to 13 sources, with a median of 8.5. This is in excellent agreement with our findings around the two sources EGS-zs8-1 and EGS-zs8-2. Both the simulations and the observations thus find a boost in the number counts around luminous sources compared to the field by $\sim3-5\times$.

\subsection{Ionized Bubble Sizes}

\begin{figure}
\centering
\includegraphics[width=\linewidth]{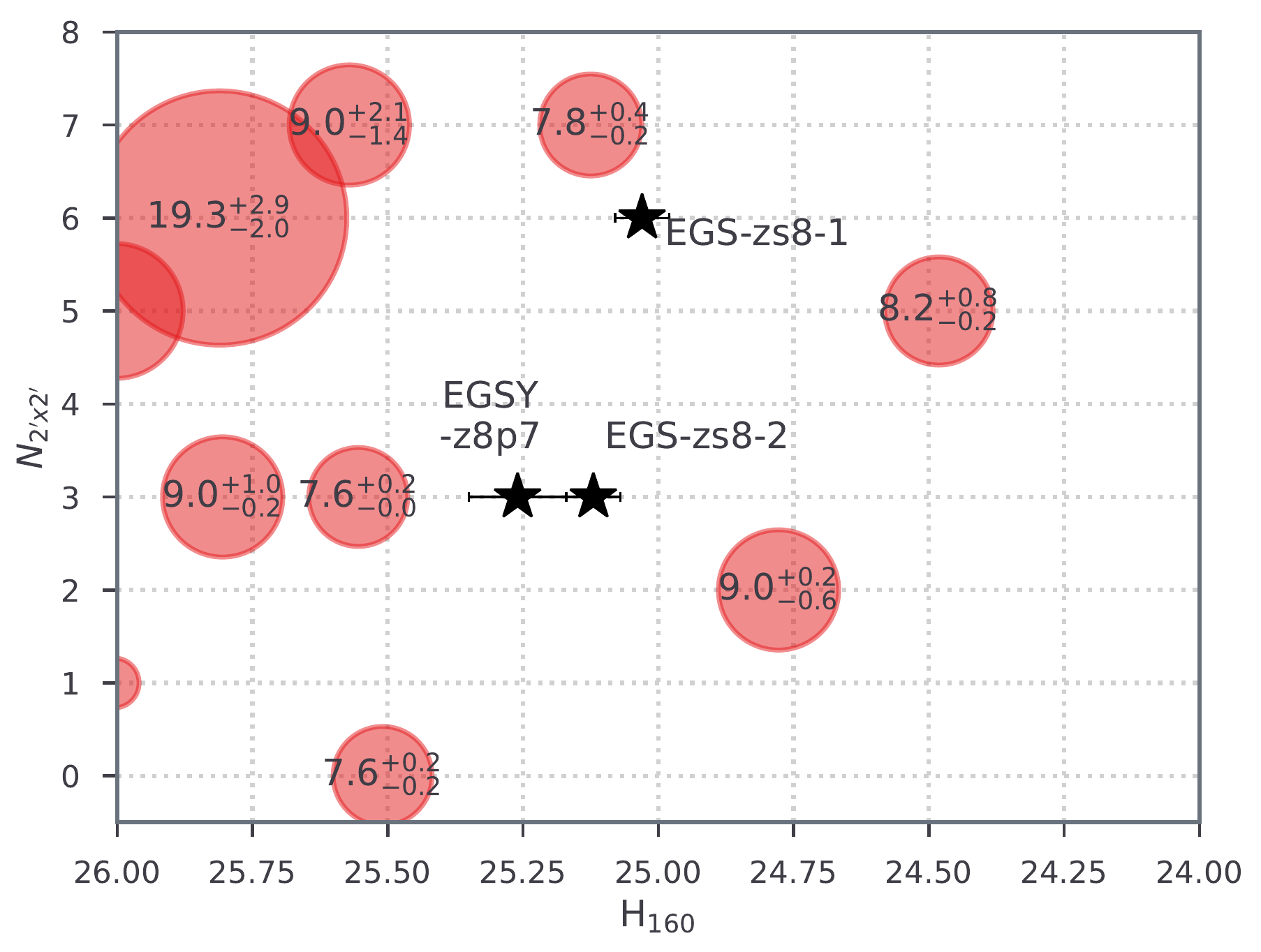}
\vspace{-12pt}
\caption{Number of neighboring galaxies brighter than ${\rm H}_{160}{=}27.5$ mag for EGS-zs8-1, EGS-zs8-2 and EGSY-z8p7 (see the star symbols) as a function of their UV magnitudes. Theoretical expectation using the DRAGONS simulation (see more in e.g., Fig. 2 of \citealt{Qin2021}) is shown with circles for comparison. The size of the corresponding \ion{H}{ii} bubbles (in units of cMpc; indicated by varying circle sizes) is shown with the uncertainties drawn from 500 mock observations.
}
\label{fig:SimBubbles}
\end{figure}

The \textsc{Meraxes} simulation further allows us to measure the radii of the ionized bubbles around galaxies. \citet{Qin2021} show that the most luminous sources sit in some of the largest ionized bubbles, resulting in a higher total transmission of the Ly$\alpha$ emission line \citep[see also][]{Wyithe2005,Geil2017,Hutter17}. The most luminous simulated central sources lie in ionized regions with measured radii of around 8-9 comoving Mpc, corresponding to $\sim1$ physical Mpc at these redshifts (see Fig. \ref{fig:SimBubbles}). We can thus expect similar ionized region sizes around our observed galaxies. Using simple estimates, this is indeed what we infer as discussed below. 

To convincingly conclude that all of the associated sources we find for
our sample lie within the respective ionised bubbles would require
both spectroscopic confirmation of the association and, ideally, the presence of
Lyman alpha emission. Nonetheless, using the same approach as in \citet{Matthee2018} we can estimate the sizes of the ionized bubbles that each of our sources individually could carve out of the neutral IGM. Assuming a spherical shape of the ionized regions around the galaxies, the expected radii range from 0.5 pMpc up to 1.1 pMpc. These results are consistent with previous estimates in the literature for possible bubble sizes at these redshifts \citep[see e.g.,][]{Castellano2018,Tilvi2020,Endsley2021}.

While \citet{Tilvi2020} have confirmed the physical association of two sources around EGS-zs8-1 through spectroscopy, the new galaxies identified in this paper only have photometric redshifts available at the moment. The respective uncertainties in 3D distances are thus too large to determine the physical associations of these galaxies with the central UV luminous targets. However, our detection of an enhancement of fainter galaxies around all these sources points at a larger scale overdensity in the EGS field \citep[see also][]{Bouwens2015, Bouwens2019, Finkelstein2021}. 

Interestingly, the two UV luminous LAEs EGS-zs8-1 and EGS-zs8-2 are only separated by 9.1 pMpc from each other. It is thus possible that these two sources lie in a larger, ionized region that spans from $z=7.5$ to $z=7.7$. Future observations, e.g., with JWST spectroscopy will be needed to confirm this.

\section{Summary and Conclusions}
\label{sec:conclusions}

In this paper, we identified faint sources at $z>7$ around three confirmed UV luminous Lyman $\alpha$ emitters and another, nearby $z_\mathrm{phot}\sim9$ target in the CANDELS/EGS field. We presented dedicated HST data and combined this with ancillary HST and Spitzer/IRAC imaging in this field. 

We find a significant enhancement of fainter galaxies within a WFC3/IR pointing of 4.5 arcmin$^2$ around each of these UV luminous, central sources. 
By comparing the number of detected galaxy candidates with the expected numbers from the UV LFs at these redshifts and with the numbers in the two GOODS fields, we estimate these areas to be enhanced by $3-9\times$ compared to the average field. 

Our observational findings are in excellent agreement with the predictions form the \textsc{Meraxes} simulation \citep{Qin2021}, which also find a boost in fainter neighbouring sources around UV luminous galaxies. In these simulations, the combined ionizing photon output of the ensemble of galaxies around the central sources produce ionized regions with radii of $\sim1$ pMpc, allowing Ly$\alpha$ photons to be transmitted even when the surrounding IGM is still highly neutral. 

Combined with these simulation results, our observational finding of overdense regions around these luminous $z=7.5-8.7$ Ly$\alpha$ emitters thus strongly suggest that the local galaxy environment plays a driving role in the Ly$\alpha$ transmission. Specifically, the field around EGSY-z8p7 at $z_\mathrm{spec}=8.683$ is enhanced by a factor $\sim8.7\times$, suggesting that this source might sit in one of the first overdensities and ionized bubbles in the Universe, only $\sim500$ Myr after the Big Bang.

The crucial next step is to obtain spectroscopic redshifts for these overdensities in order to map out their 3D structure and estimate the size of the overall ionized regions. The approved JWST cycle 1 program GO-2279 \citep{NaiduJWSTprop} will achieve exactly this based on NIRCam/grism observations of EGSY-z8p7 and EGS910-3, which is only separated by 3.7 arcmin. These observations will provide a sample of [OIII]5007 emission line selected sources at $z=7-9$, and will thus significantly improve the overdensity measurement. Other NIRCam/grism and NIRSpec observations over other fields are poised to identify a large number of ionized bubbles in the EoR in the near future. This will  further clarify the importance of other possible origins for enhanced Ly$\alpha$ transmission from UV luminous galaxies, such as velocity offsets or very hard and strong ionization fields, e.g., from an AGN contribution.

The fact that we find an overdensity around each of the three UV luminous Ly$\alpha$ emitting galaxies only based on photometric redshifts strongly suggests that the Ly$\alpha$ transmission is intimately connected to the overall galaxy environment that helps to create ionized regions. Turning this around, these UV luminous, early Ly$\alpha$ emitters can be used to pinpoint the earliest overdense regions in the Universe, where star-formation and reionization first started. This will become particularly interesting with the availability of larger samples of UV luminous LBGs and LAEs from future wide-area surveys with the Euclid or Roman Space Telescopes and follow-up spectroscopy.

\section*{Acknowledgements}
We acknowledge support from: the Swiss National Science Foundation through the SNSF Professorship grant 190079 (PAO, JK).
The Cosmic Dawn Center (DAWN) is funded by the Danish National Research Foundation under grant No.\ 140.
This research was supported by the Australian Research Council Centre of Excellence for All Sky Astrophysics in 3 Dimensions (ASTRO 3D), through project number CE170100013. Part of this work was performed on the OzSTAR national facility at Swinburne University of Technology, which is funded by Swinburne University of Technology and the National Collaborative Research Infrastructure Strategy (NCRIS).
ST is supported by the 2021 Research Fund 1.210134.01 of UNIST (Ulsan National Institute of Science \& Technology).
RSE acknowledges funding from the European Research Council (ERC) 
under the European Union's Horizon 2020 research and innovation programme 
(grant agreement No 669253).
RJB and MS acknowledge support from TOP grant TOP1.16.057.
RS acknowledges support from a STFC Ernest Rutherford Fellowship (ST/S004831/1).
AH acknowledges support from the European Research Council's starting grant ERC StG-717001 (``DELPHI'').
AZ acknowledges support by Grant No. 2020750 from the United States-Israel Binational Science Foundation (BSF) and Grant No. 2109066 from the United States National Science Foundation (NSF), and by the Ministry of Science \& Technology, Israel.

\section*{Data Availability}
The HST data underlying this article will be publicly released as High-Level Science Product (HLSP) as part of the Hubble Legacy Fields (HLF) project \citep{Illingworth16,Whitaker19}. A pre-release version can be shared on reasonable request to the corresponding author.


\bibliographystyle{mnras}
\bibliography{EGSBubblePaper} 



\appendix

\section{Stamps and Photometric Redshift Likelihoods for the High-z Sample}

\begin{figure*}
\centering
\includegraphics[width=0.45\linewidth]{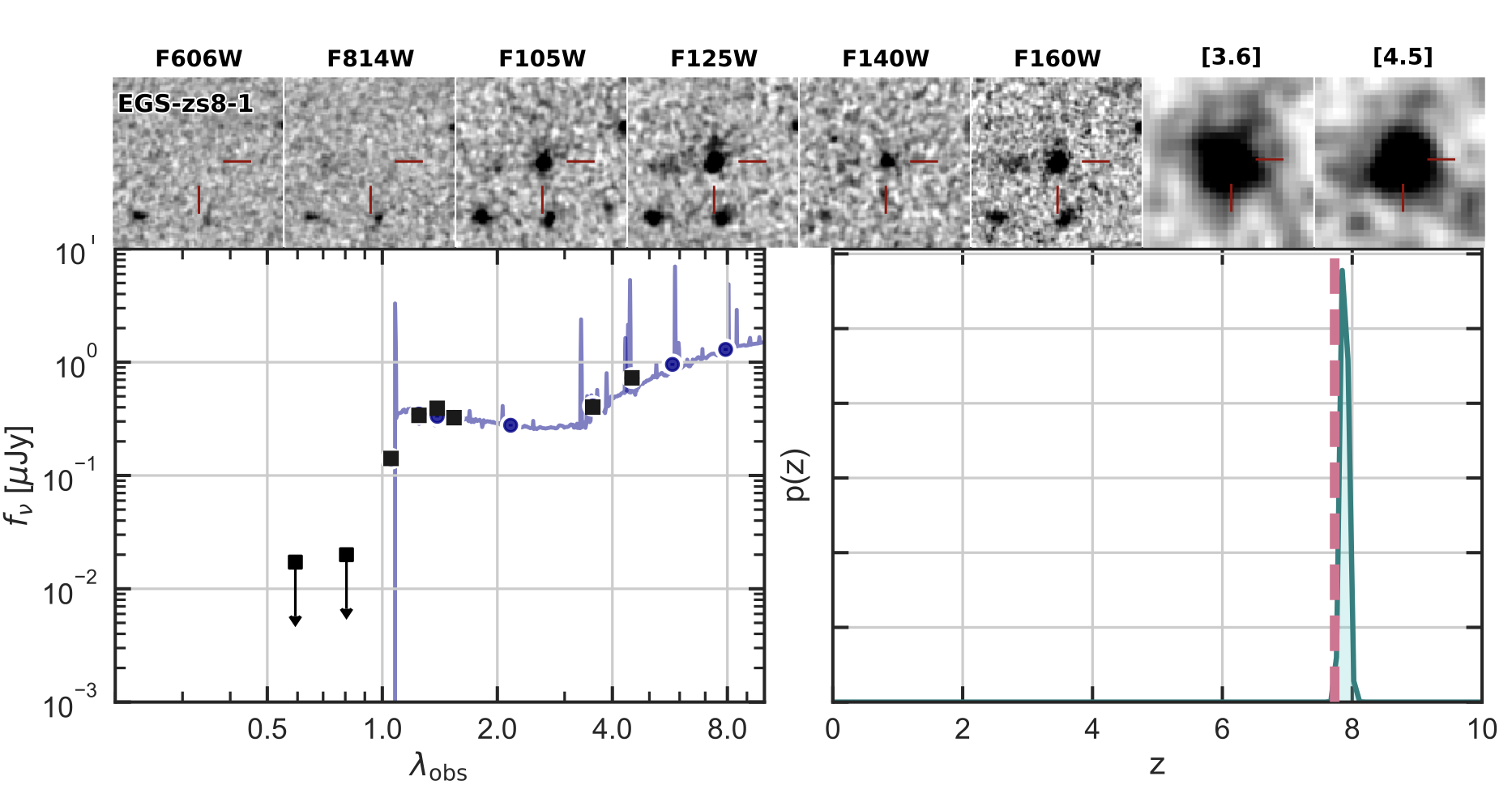}
\quad
\includegraphics[width=0.45\linewidth]{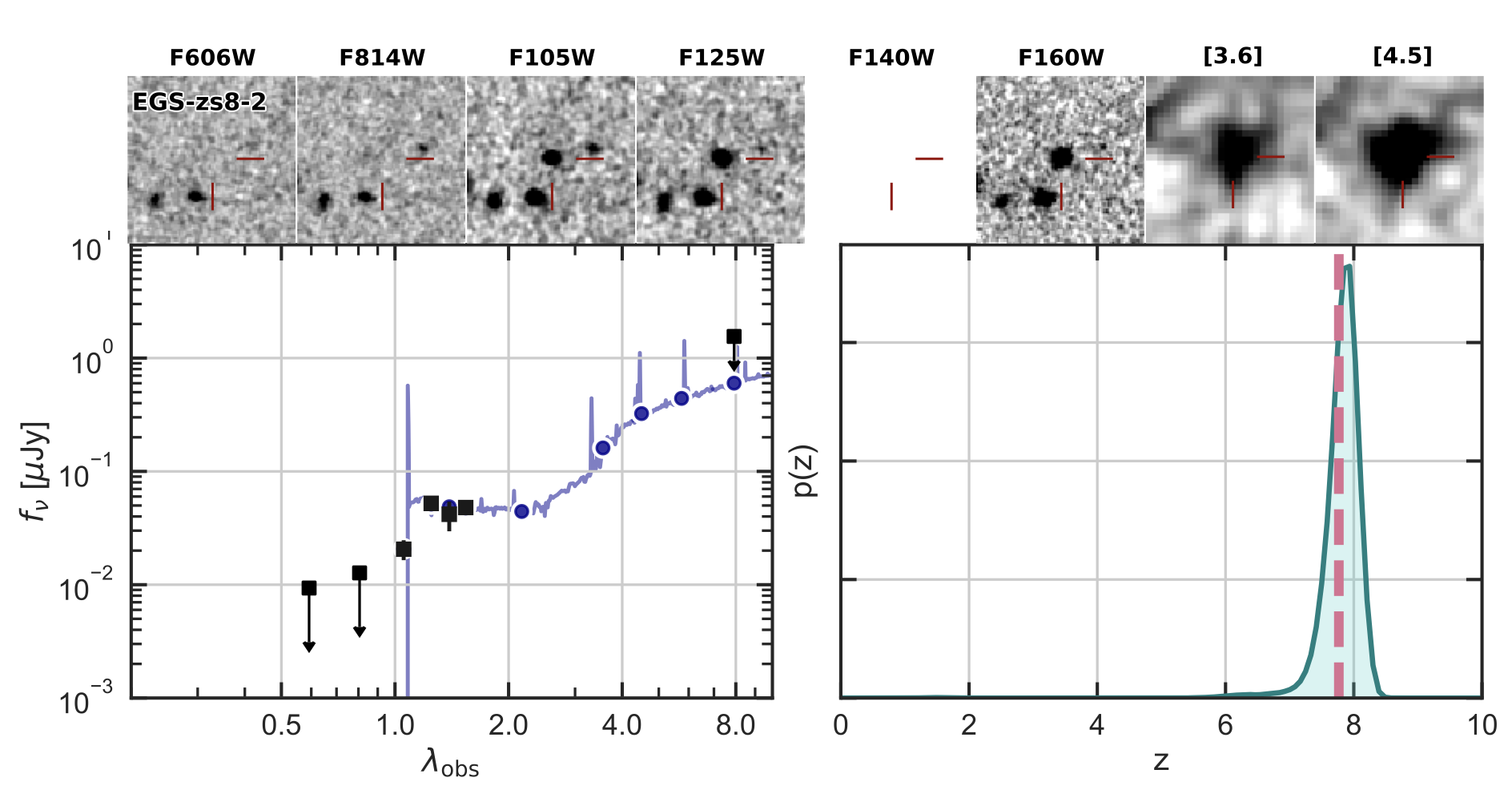}
\medskip
\includegraphics[width=0.45\linewidth]{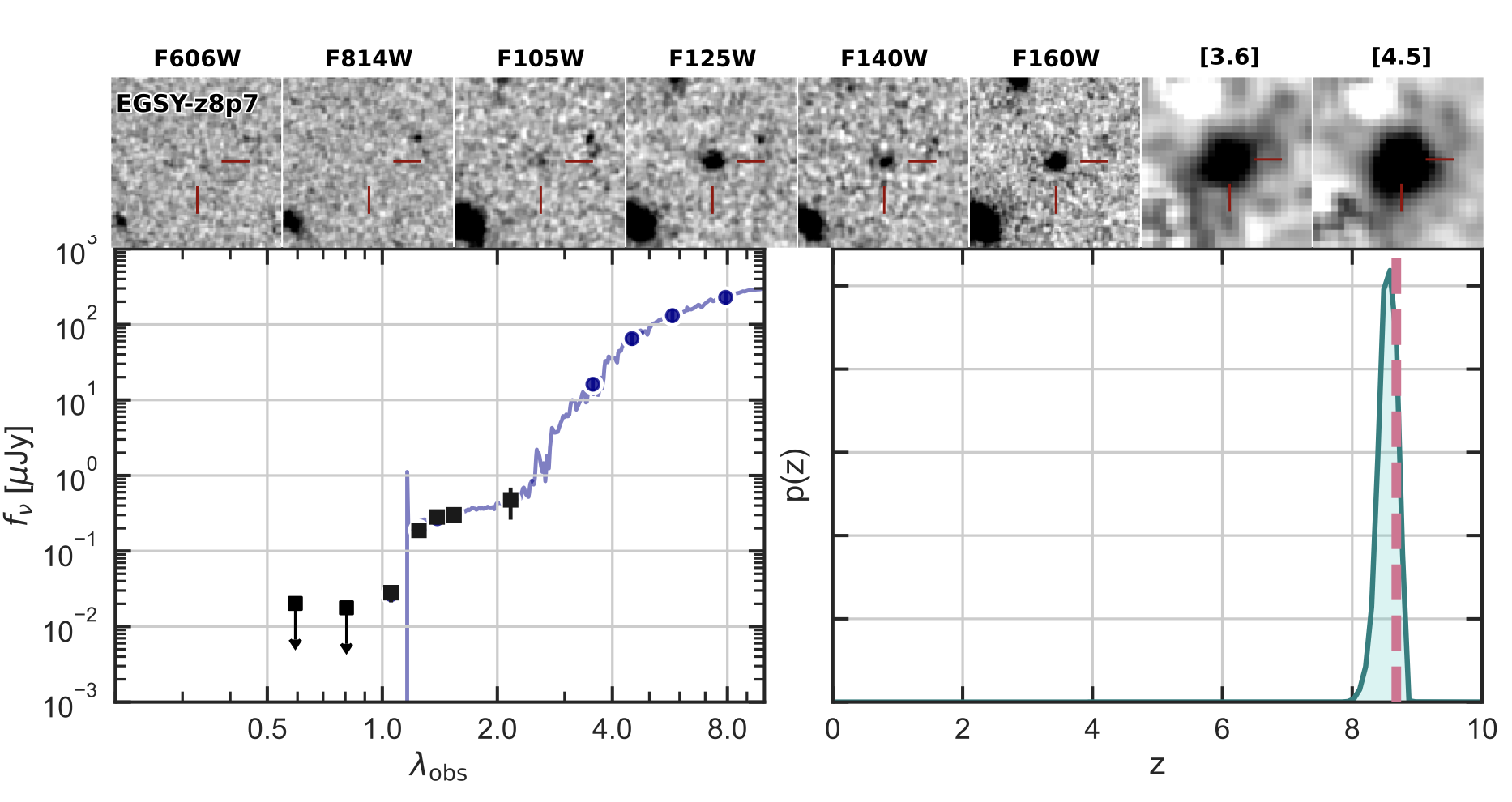}
\quad
\includegraphics[width=0.45\linewidth]{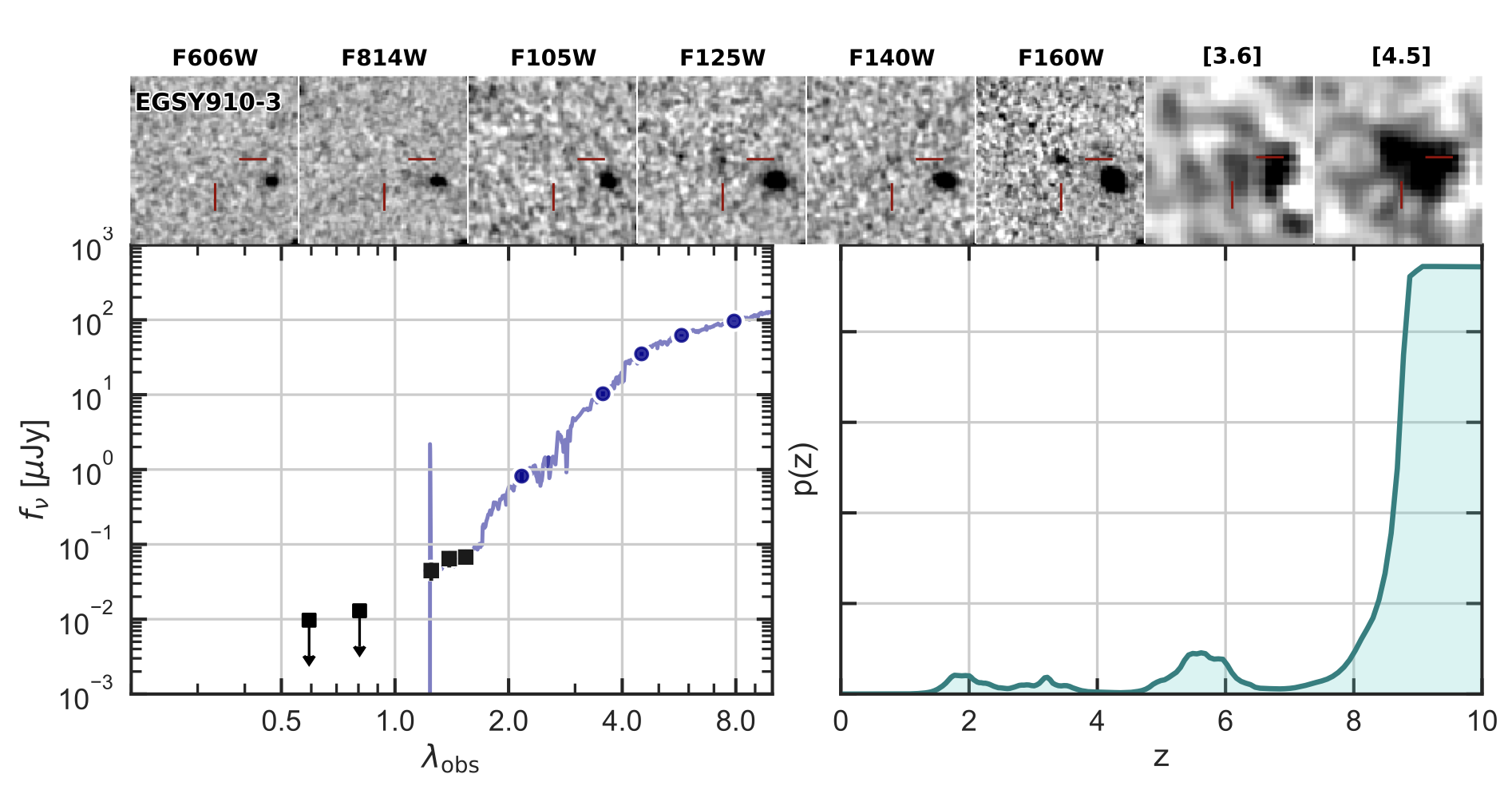}

\caption{Multi-wavelength images and SEDs of the UV-luminous central targets: three Lyman $\alpha$ emitters and EGS910-3. For each source, the top panels show, from left to right, HST/ACS $V_{606},\ I_{814}$, HST/WFC3 $Y_{105},\ J_{125},\ H_{160}$, and Spitzer/IRAC $3.6\mu m$ and $4.5\mu m$ stamps. The lower panels show the SED (purple curve with uncertainties), the photometry (black squares with uncertainties), and the resulting probability distribution of redshift $\mathcal{P}(z)$ (green curve). The pink vertical lines represent the spectroscopic redshift of the Ly$\alpha$ emitters, showing that the photometric redshifts are accurate for these sources. The last source, EGS910-3, does not have a spectroscopic redshift measurement yet. Our photometric redshift is in excellent agreement with the original determination of \citet{Bouwens2016d}.}
\label{fig:sed}
\end{figure*}

\begin{figure*}
\centering
\includegraphics[width=\linewidth]{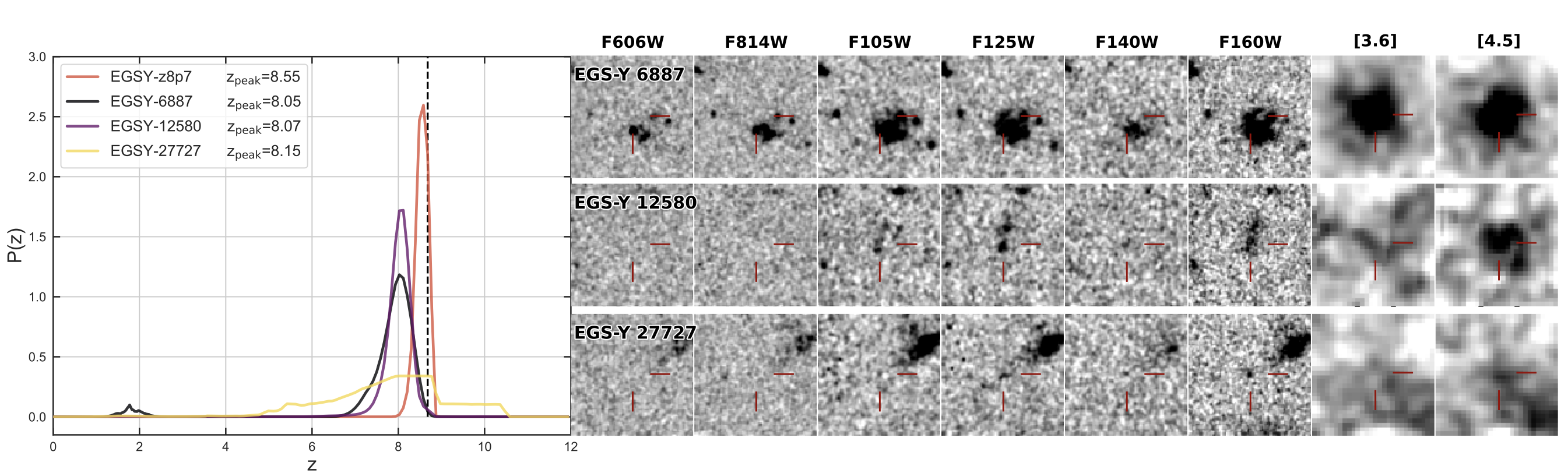}

\caption{Redshift probability distribution functions $\mathcal{P}(z)$ and stamps of fainter $z_\mathrm{phot}>8$ sources in the pointing around EGSY-z8p7. The $\mathcal{P}(z)$ of the UV luminous central target is also shown (orange line), indicating that the neighbouring sources have a significant probability to be physically associated to the central source.
The stamps show the same filters as \ref{fig:sed} and are labelled on the top.}
\label{fig:egsz8p7group}
\end{figure*}

\begin{figure*}
\centering
\includegraphics[width=\linewidth]{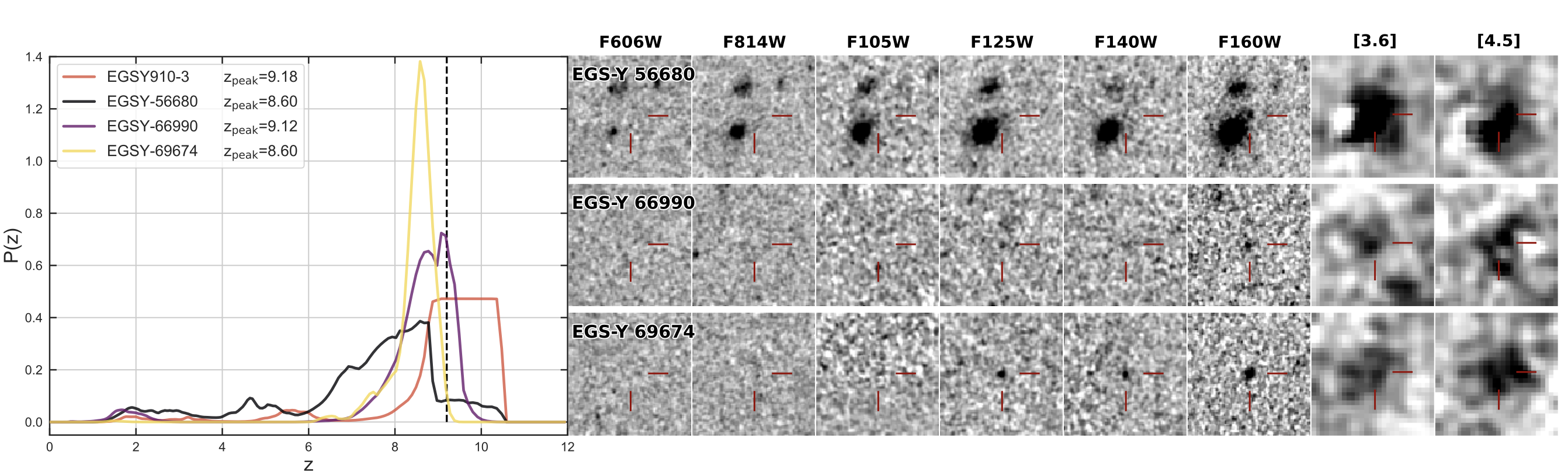}
\caption{The same as Figure \ref{fig:egsz8p7group} for the neighbouring sources around EGS910-3 with $z_\mathrm{phot}>8$.}
\label{fig:egsz9group}
\end{figure*}

\begin{figure*}
\centering
\includegraphics[width=\linewidth]{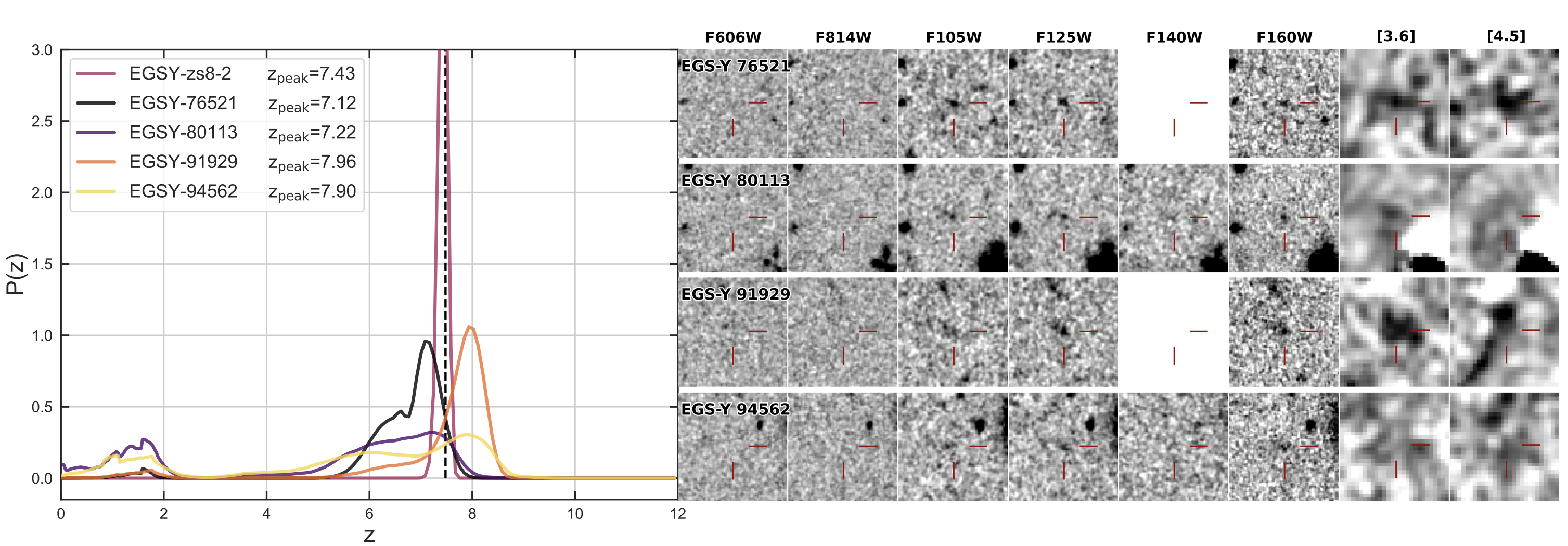}
\caption{The same as Figure \ref{fig:egsz8p7group} for the neighbouring sources around EGS-zs8-2 with $z_\mathrm{phot}=7-8$.}
\label{fig:egszs82group}
\end{figure*}

\begin{figure*}
\centering
\includegraphics[width=\linewidth]{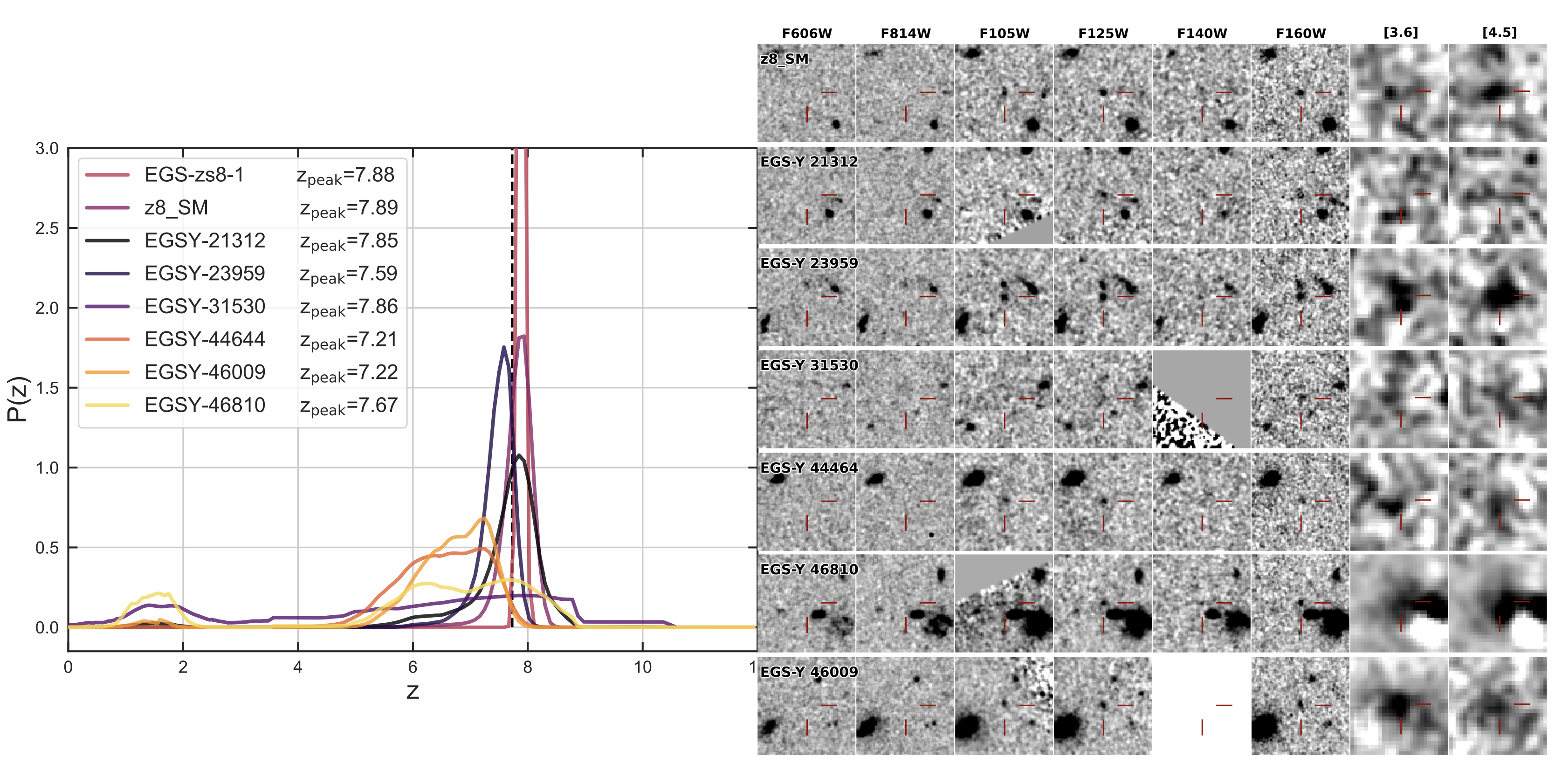}
\caption{The same as Figure \ref{fig:egsz8p7group} for the neighbouring sources around EGS-zs8-1 with $z_\mathrm{phot}=7-8$.}
\label{fig:egszs81group}
\end{figure*}

Here, we show the image stamps, SEDs and photometric redshift likelihood functions for all sources that were selected in our fields. In \ref{fig:sed}, we show how our photometric redshifts compare to the spectroscopic redshifts measured for the UV-luminous central targets.
The neighbouring sources around the four central targets are show in Fig. \ref{fig:egsz8p7group}, \ref{fig:egsz9group}, \ref{fig:egszs82group}, and \ref{fig:egszs81group}. 
While there is clearly a relatively large spread, the combined likelihood functions for the neighbours agree well with the redshifts of the central sources, indicating that the majority of our selected galaxies indeed have a good chance to be physically associated with the central UV-luminous sources. Spectroscopy will be needed to confirm this in the future, however.

\bsp	
\label{lastpage}
\end{document}